\documentclass[
 reprint,
 superscriptaddress,
 aps,
 prl,
]{revtex4-2}

\usepackage{amsmath,amsthm,amsfonts,amssymb,amscd}
\usepackage{graphicx}
\usepackage{dcolumn}
\usepackage{bm}
\usepackage{slashed}
\usepackage{xcolor}

\begin{document}

\title{Large-Width New Physics at Colliders: A Gauge-Invariant Resummation Approach}

\author{Yin-Fa Shen}
\email{yin-fa.shen@Vanderbilt.Edu}
\author{Alfredo Gurrola}
\email{alfredo.gurrola@Vanderbilt.Edu}
\affiliation{Department of Physics and Astronomy, Vanderbilt University, Nashville, TN, 37235, USA}

\begin{abstract}
    Broad resonances challenge the standard Monte-Carlo treatment of unstable particles, which introduces a Breit–Wigner width into leading-order matrix elements and can generate unphysical gauge artifacts. We develop a gauge-consistent framework that combines a Dyson-resummed propagator with Slavnov–Taylor-identity–implied resummed vertices, enabling a consistent implementation in \textsc{MadGraph5}. In the Type-I seesaw model, heavy Majorana neutrinos naturally satisfy $\Gamma \sim m$, leading to strong departures from the Breit–Wigner lineshape, distorted angular correlations, and significant modifications to both $s$- and $t$-channel dynamics. Comparing with the normal and complex-mass schemes, we find that standard treatments can substantially misestimate cross sections and kinematic distributions in the large-width regime. Our results show that existing collider limits on heavy neutrinos—and, more generally, on any broad resonance—should be revisited within a fully resummed framework, opening new opportunities for both experimental searches and theoretical model building.
\end{abstract}

\maketitle

\textit{Introduction}---Searches for new physics (NP) remain a central objective of current and future high-energy collider programs. Many NP frameworks predict new unstable particles that decay into Standard Model (SM) states with distinctive kinematic signatures. While numerous analyses target narrow resonances, a broad class of NP models predict states with intrinsically large decay widths. Accurately modeling large-width effects is therefore essential for robust reconstruction of NP signals.\\
\indent In perturbation theory, decay widths originate from loop corrections via Dyson resummation of one-particle-irreducible (1PI) self-energies~\cite{Dyson:1949ha, Schwinger:1951ex, Schwinger:1951hq, Veltman:1963th} in the \emph{dressed} propagator. A consistent treatment thus requires next-to-leading-order (NLO) calculations, which are often computationally demanding. Consequently, Monte-Carlo event generators such as \textsc{MadGraph5}~\cite{Alwall:2014hca, Frederix:2018nkq} typically insert finite widths into leading-order (LO) propagators through a Breit–Wigner form and rely on the narrow-width approximation (NWA). This ad-hoc combination of an NLO propagator with LO amplitudes formally breaks gauge invariance; however, the resulting gauge artifacts scale as $\mathcal{O}(\Gamma/m)$~\cite{Uhlemann:2008pm} and are negligible for narrow SM resonances.\\
\indent The issue becomes more subtle in NP scenarios where the Higgs or gauge sectors are extended and mass–coupling relations differ from the SM. Representative examples include heavy Majorana neutrinos in seesaw models~\cite{Minkowski:1977sc, Gell-Mann:1979vob, Yanagida:1979as, Mohapatra:1979ia} and theories with additional $Z^\prime$ or $W^\prime$ gauge bosons~\cite{Pati:1974yy, Mohapatra:1974hk, Senjanovic:1975rk}. Experimental analyses often treat masses and couplings as independent to maintain model independence, which naturally scans over regions with $\Gamma \sim m$, where the NWA fails even though the decay width formally originates at NLO. In such regions, gauge-dependent artifacts can become sizable~\cite{Stuart:1991xk, Sirlin:1991fd}, resonance lineshapes deviate strongly from a Breit–Wigner form~\cite{Goria:2011wa, Kauer:2012hd}, and overlap with SM backgrounds can increase. A gauge-consistent and physically meaningful treatment of large width effects is therefore crucial.\\
\indent \textsc{MadGraph5} employs the \emph{complex-mass scheme}, which defines propagator poles in the complex plane and preserves gauge invariance at LO for arbitrary widths~\cite{Denner:1999gp, Denner:2005fg}. However, this method introduces widths only through analytic continuation and does not capture the full dynamical structure of intrinsically NLO large-width effects. While propagator resummation is well established, vertex resummation is governed by the Bethe–Salpeter equation (BSE)~\cite{Salpeter:1951sz}, which is difficult to solve and thus absent from automated NLO tools.\\
\indent Most NP models are gauge invariant, and after gauge fixing, the BRST symmetry~\cite{Becchi:1975nq, Tyutin:1975qk} ensures that physical amplitudes remain gauge independent through the Slavnov–Taylor identities (STIs)~\cite{Taylor:1971ff, Slavnov:1972fg}. These identities relate propagators and vertices, implying that once a propagator is resummed, the associated vertices can, in principle, be reconstructed without solving the full BSE.\\
\indent In this Letter, we examine the limitations of current treatments in NP scenarios characterized by large decay widths. We show that standard approaches can yield misleading predictions for both total rates and differential observables, revealing that \emph{large-width effects induce qualitative changes that have long been overlooked} in collider NP searches. Using the Type-I seesaw model as a concrete example, \emph{we develop an approach that consistently incorporates the interplay among large-width effects, interference, and kinematics}, improving simulation accuracy beyond the NWA and LO complex-mass approximations. We derive the STI-implied resummed vertices corresponding to a propagator modified by appropriate self-energies and implement them in \textsc{MadGraph5}. Our results suggest that previous heavy neutrino searches~\cite{CMS:2018jxx, CMS:2022hvh, CMS:2024xdq, ATLAS:2024rzi}—and, more broadly, searches for wide resonances—may warrant reinterpretation within a fully resummed framework.\\
\indent \textit{Benchmark Model}---A minimal and well-motivated explanation for tiny neutrino masses is the introduction of heavy Majorana neutrinos, whose large mass terms generate light neutrino masses through the seesaw mechanism~\cite{Minkowski:1977sc, Gell-Mann:1979vob, Yanagida:1979as, Mohapatra:1979ia}. For simplicity, we consider a right-handed singlet neutrino. After electroweak symmetry breaking, its interactions with SM fields are
\begin{equation}\label{HNint}
\begin{split}
\mathcal{L} \ni \sum_{\ell} \Big[
&-\frac{g\,V^{*}_{\ell N}}{\sqrt{2}}\;
   \bar N \slashed{W}^{+} P_L \ell -\frac{g\,V^{*}_{\ell N}}{2\cos\theta_W}\;
   \bar N \slashed{Z} P_L \nu_\ell \\
&-\frac{g\,m_N\,V^{*}_{\ell N}}{\sqrt{2} m_W}\;
   \bar N \tilde\Phi^\dagger L_\ell 
+ \text{h.c.} \Big],
\end{split}
\end{equation}
where $V_{\ell N}$ is the active–sterile mixing parameter, $L_{\ell}$ denotes the left-handed lepton doublet, and $\tilde{\Phi}$ the conjugate Higgs doublet. From Eq.~\eqref{HNint} and the Goldstone Boson Equivalence Theorem~\cite{Cornwall:1974km, Vayonakis:1976vz}, the heavy neutrino decay width scales as
\begin{equation}
    \Gamma_N \sim  \left|V_{\ell N}\right|^2\left(\frac{m_N}{m_W}\right)^2 m_N.
\end{equation}
Thus $\Gamma_N$ grows cubically with $m_N$ and linearly with $|V_{\ell N}|^2$. For TeV-scale masses and sizable mixing, $\Gamma_N$ can become comparable to or even exceed $m_N$, invalidating the NWA and signaling a breakdown of perturbative unitarity~\cite{Lee:1977eg}. Nevertheless, current LHC searches typically assume a Breit–Wigner lineshape across the entire parameter space~\cite{CMS:2018jxx, CMS:2022hvh, CMS:2024xdq, ATLAS:2024rzi}, even in regimes where this approximation is not theoretically justified.\\
\indent \textit{Methodology}---To capture large-width effects, we work in the Feynman gauge and resum the heavy neutrino propagator,
\begin{equation}
    \frac{i}{\slashed{p} - m_\text{R} + \Sigma_\text{R}(\slashed{p})},
\end{equation}
where the renormalized self-energy $\Sigma_\text{R}(\slashed{p})$ is defined such that the pole lies at $m_\text{R} = m_N - i\Gamma_N/2$, preserving gauge invariance of the pole position~\cite{Gambino:1999ai}. Neglecting SM lepton masses, assuming mixing with a single generation, and retaining only the leading large-$m_N$ terms, we obtain
\begin{equation}\label{ferpro}
    \frac{i}{\slashed{p} - m_\text{R} + \Sigma_\text{R}(\slashed{p})} = \frac{i}{A\,\slashed{p} - B\, m_\text{R}} = \frac{i(A\,\slashed{p} + B\, m_\text{R})}{A^2 p^2 - B^2 m^2_\text{R}},
\end{equation}
with
\begin{subequations}\label{AB}
\begin{align}
A &= 1 + \frac{\Gamma_N}{2\pi m_N}
       \sum_i \mathrm{BR}_i
       \Bigg[
         \frac{2 m_R^{2}}{m_R^{2}-m_i^{2}}
         - \log\!\left(
             \frac{|p^{2}-m_i^{2}|}{m_R^{2}-m_i^{2}}
           \right)
\nonumber \\
&\hphantom{=1 + \frac{\Gamma_N}{2\pi m_N}\sum_i\mathrm{BR}_i\Bigg[}
         +\, i\pi\,\theta(p^{2}-m_i^{2})
       \Bigg], \\
B &= 1 + \frac{\Gamma_N}{2\pi m_N}
       \sum_i \mathrm{BR}_i
       \left(
         \frac{2 m_R^{2}}{m_R^{2}-m_i^{2}}
       \right),
\end{align}
\end{subequations}
where the sum runs over $i=W,Z,h$, $\mathrm{BR}_i$ is the corresponding branching ratio, and $\theta(x)$ is the unit step function. In our numerical studies, we take $\mathrm{BR}(N \to W^\pm\ell^\mp)=50\%$ and $\mathrm{BR}(N\to Z\nu_\ell)=\mathrm{BR}(N\to h\nu_\ell)=25\%$.

The unit step function indicates that an imaginary part appears only once $p^2$ exceeds the threshold of a given channel. This differs from the \textsc{MadGraph5} prescription~\cite{Alwall:2014hca, Frederix:2018nkq}, which inserts a Breit–Wigner width and \emph{sets it to zero for $t$-channel exchanges by default}. As Eq.~\eqref{AB} shows, even when the imaginary part vanishes in the $t$-channel, real contributions of comparable size remain. Conversely, the complex-mass scheme embeds the imaginary term directly into the mass definition, keeping it nonzero even in $t$-channel kinematics. Thus, at LO, neither approach fully captures the dynamics associated with large widths.

Moreover, inserting the self-energy into the propagator through Dyson resummation can still violate gauge invariance unless the corresponding vertex corrections are consistently resummed. In principle, these vertex resummations are encoded in the Bethe–Salpeter equation~\cite{Salpeter:1951sz}, a nonlinear integral equation that is notoriously difficult to solve. In the present setup, however, the introduced right-handed neutrino is a gauge singlet, so gauge invariance of the Lagrangian is preserved. After gauge fixing, BRST symmetry~\cite{Becchi:1975nq, Tyutin:1975qk} implies a set of Slavnov–Taylor identities (STIs)~\cite{Taylor:1971ff, Slavnov:1972fg} that relate Green’s functions and guarantee the gauge independence of physical amplitudes. For the heavy neutrino–gauge boson–SM fermion vertex $\Gamma^\mu$, the relevant STI reads
\begin{equation}\label{STI}
    k^\mu \Gamma_\mu = m_i\times i\Gamma_\phi + i [S^{-1}_N(\slashed p)\cdot P_L - P_R\cdot S^{-1}_f(\slashed q)],
\end{equation}
where $\Gamma_\phi$ is the corresponding heavy neutrino–Goldstone–SM fermion vertex from the Yukawa sector, and $p$, $k$, and $q$ are the momenta of the outgoing heavy neutrino, gauge boson, and incoming SM fermion, respectively. Here $m_i$ is the gauge-boson mass, $P_{L/R}$ are chiral projectors, and $S^{-1}$ denotes the inverse fermion propagator. In a nonchiral Abelian theory without spontaneous symmetry breaking ({\it e.g.}, QED), both $\Gamma_\phi$ and the projectors vanish, recovering the Ward–Takahashi identity~\cite{Ward:1950xp, Takahashi:1957xn}.

Because STIs relate to amputated Green’s functions, explicit gauge couplings are omitted in Eq.~\eqref{STI}. For scalar (Yukawa-like) vertices such as $\Gamma_\phi$ or $\Gamma_h$, the couplings must be retained to maintain dimensional consistency. BRST symmetry guarantees that the STIs hold order by order, enabling the consistent reconstruction of resummed vertices once the propagator $S(\slashed{p})$ is resummed.

These resummed vertices correspond to the Ball–Chiu construction~\cite{Ball:1980ay, Ball:1980ax}, in which $\Gamma^\mu$ is expanded in a Clifford basis with coefficients fixed by STIs and physical constraints. For the heavy neutrino case, the vertices take the form
\begin{subequations}\label{vertex}
\begin{align}
&\Gamma^\mu = \left[\frac{A_N + A_f}{2}\gamma^\mu + \frac{A_N - A_f}{2}\frac{p^\mu + q^\mu}{p^2 - q^2}(\slashed p + \slashed q)\right] P_L, \\
&\Gamma_\phi = \frac{1}{m_i} (m_\text{R} B_N \cdot P_L - P_R \cdot B_f m_f ),\\
&\Gamma_h = \frac{1}{m_W} (m_\text{R} B_N \cdot P_L),
\end{align}
\end{subequations}
where the coefficients $A$ and $B$ are those defined in Eq.~\eqref{ferpro}. Physically, $A$ encodes wave-function corrections entering gauge-boson vertices, while $B$ modifies fermion mass terms and thereby mass-related couplings. In the limit $A, B\to 1$, these expressions smoothly reduce to their tree-level forms.\\
\indent \textit{Collider Implications}---We consider three computational schemes: the normal scheme, the complex-mass scheme, and our resummed scheme. For the normal and complex-mass schemes, we follow the default \textsc{MadGraph5}~\cite{Alwall:2014hca, Frederix:2018nkq} implementation, including its built-in Breit–Wigner treatment of unstable particles. In contrast, the resummed scheme requires explicit modifications to the model files, enabled by the flexibility of the Universal FeynRules Output (UFO) format~\cite{Darme:2023jdn}. We implement the resummed propagators and vertices, disable the default \textsc{MadGraph5} setting that enforces vanishing widths for $t$-channel propagators, and switch to the Feynman gauge for consistency. The coefficients $A$ and $B$ for all SM fermions are fixed to unity, preserving SM interactions at LO, while the heavy neutrino corrections are included at NLO to capture large-width effects.

\begin{figure}
    \centering
    \includegraphics[width=0.43\linewidth]{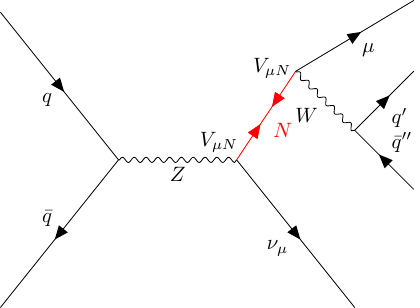}
    \hspace{0.1\linewidth}
    \includegraphics[width=0.32\linewidth]{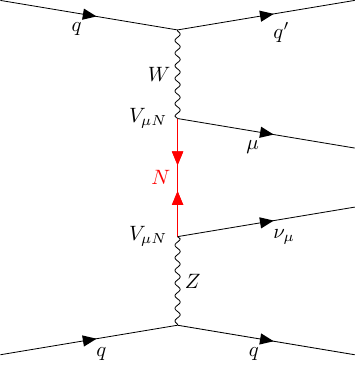}
    \caption{Representative diagrams for $pp \to \mu \nu_\mu jj$ arising from the $s$-channel (left) and the $t$-channel (right) processes.}
    \label{fig:Feyn}
\end{figure}

We simulate the electroweak process $pp \to \mu \nu_\mu j j$ at $\sqrt{s}=13\,\mathrm{TeV}$, incorporating all interactions in Eq.~\eqref{HNint}. As shown in Fig.~\ref{fig:Feyn}, production receives contributions from both $s$-channel and $t$-channel diagrams. We also simulate the same-sign dilepton channel $pp\to\mu^{\pm}\mu^{\pm} jj$ (likewise pure electroweak), directly relevant to existing searches~\cite{CMS:2018jxx, CMS:2022hvh, ATLAS:2024rzi} and originally proposed in Ref.~\cite{Fuks:2020att}. The motivation for studying $pp\to\mu^{\pm}\mu^{\pm} jj$ is twofold: $(i)$ to confirm our implementation reproduces known results, and $(ii)$ to provide a benchmark for comparing large-width effects across schemes. As noted, setting $V_{\ell N}=1$ yields $\Gamma_N/m_N\sim 70\%$ at $m_N=750\,\mathrm{GeV}$ and $\Gamma_N \gtrsim m_N$ for $m_N\sim\mathcal{O}(1\,\mathrm{TeV})$, signaling a breakdown of perturbation theory. Nevertheless, because several experimental analyses~\cite{CMS:2022hvh} adopt $V_{\ell N}=1$ and subsequently rescale sensitivities, we keep $V_{\ell N}=1$ in the normal and complex-mass schemes. In the resummed scheme, we instead scale $V_{\ell N}$ to enforce $\Gamma_N \leq m_N$, maintaining perturbative control at multi-TeV masses for a meaningful comparison. Focusing on the large-width regime, we scan heavy-neutrino masses in the range $750$--$10000\,\mathrm{GeV}$.

\begin{figure}
    \includegraphics[keepaspectratio, width=1\linewidth]{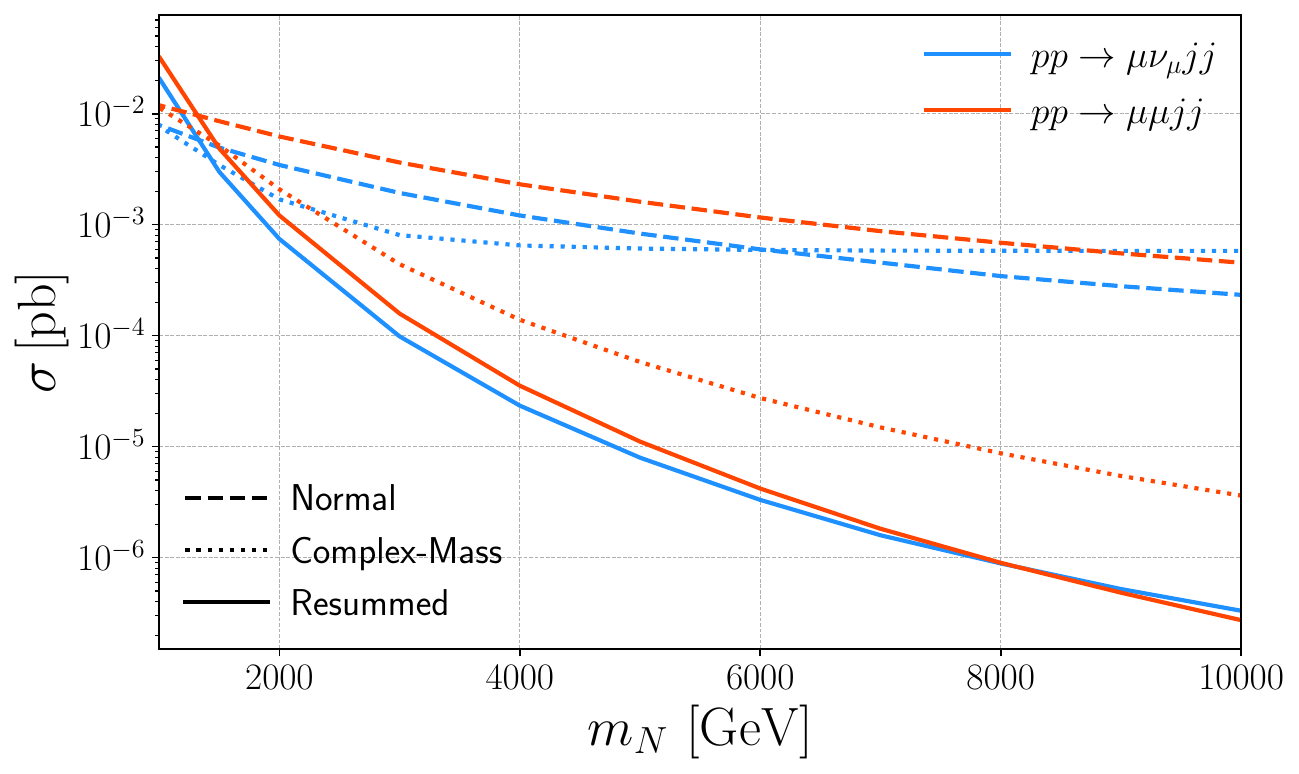}
    \caption{Cross sections for $pp\to\mu\nu_\mu jj$ and $pp\to\mu\mu jj$ versus $m_N$ in the normal, complex-mass, and resummed schemes.}
    \label{fig:cross_sections}
\end{figure}

We first verify that we reproduce the results of Ref.~\cite{Fuks:2020att}. For $m_{N}= 1.5$ $(5)\,\mathrm{TeV}$ and $V_{\ell N}=1$, we obtain $\sigma(pp\to\mu^{\pm}\mu^{\pm} jj) = 8.5$ $(1.6)\,\mathrm{fb}$ using the normal scheme, consistent with Ref.~\cite{Fuks:2020att}. The three schemes agree well at smaller masses, reflecting the effective restoration of standard treatments when $\Gamma_N/m_N$ is modest. For example, for $m_{N} = 500\,\mathrm{GeV}$ and $V_{\ell N}=1$ ($\Gamma_N/m_N \sim 30\%$), we find $\sigma(pp\to\mu^{\pm}\mu^{\pm} jj) \sim 20\,\mathrm{fb}$ in both the normal and complex-mass schemes, compared to $24\,\mathrm{fb}$ in the resummed scheme. Figure~\ref{fig:cross_sections} shows the production cross sections as functions of $m_N$ in the large-width regime. At higher masses, the normal-scheme prediction decreases smoothly with $m_N$ because its dominant $t$-channel contribution is unaffected by the large width—\textsc{MadGraph5} sets all $t$-channel widths to zero by default. In the complex-mass scheme, the width enters the definition of the complex Yukawa coupling, producing an unphysical enhancement in the Higgs-exchange amplitude: the $pp\to\mu\nu_\mu jj$ rate becomes artificially flat, while the same-sign dilepton channel—without Higgs exchange—remains highly suppressed. The resummed scheme avoids both issues: it incorporates the physical width consistently across all channels and yields cross sections that decrease monotonically with $m_N$, reflecting the expected suppression once $\Gamma_N$ becomes large.

\begin{figure}
    \includegraphics[keepaspectratio, width=1\linewidth]{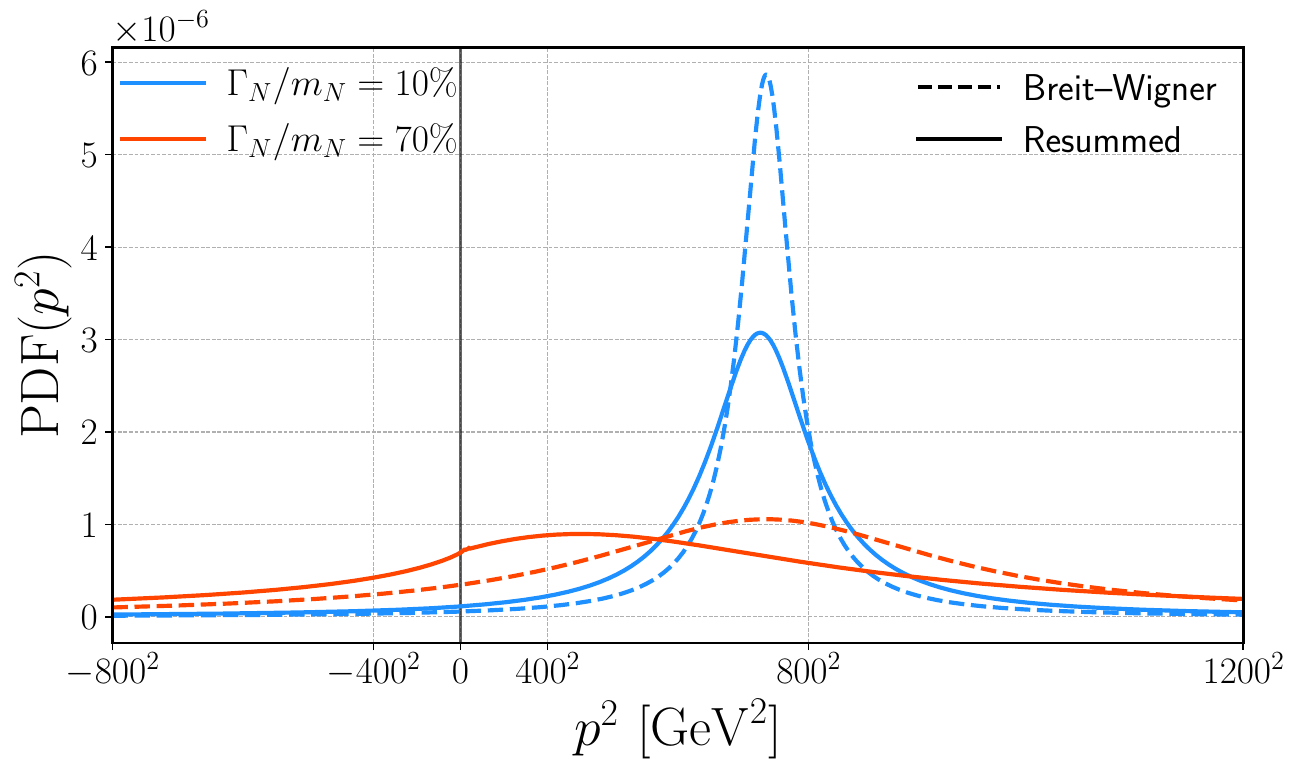}
    \caption{Normalized probability densities obtained from the squared propagator denominators for a heavy neutrino with $m_N=750\,\mathrm{GeV}$, comparing the Breit–Wigner and fully resummed lineshapes.}
    \label{fig:shape}
\end{figure}

We next compare the invariant-mass distributions implied by the Breit–Wigner and resummed propagators. Since the heavy neutrino propagator carries spinor structure, we use the squared denominators as probability density functions (PDFs). We illustrate the difference for a benchmark mass of $m_N = 750\,\mathrm{GeV}$ with two representative widths, $\Gamma_N/m_N=10\%$ and $70\%$, both within the range probed experimentally in Refs.~\cite{CMS:2018jxx, CMS:2022hvh, CMS:2024xdq, ATLAS:2024rzi}. FIG.~\ref{fig:shape} shows the resummed distribution already departs noticeably from the Breit–Wigner shape at $\Gamma_N/m_N=10\%$. For a larger width, $\Gamma_N/m_N=70\%$, the deviation becomes substantial: the peak shifts away from the physical mass, the lineshape broadens dramatically, and the space-like region contributes non-negligibly. These features highlight the breakdown of the Breit–Wigner form in the large-width regime and the need for a fully resummed treatment.

\begin{figure}
    \includegraphics[keepaspectratio, width=1\linewidth]{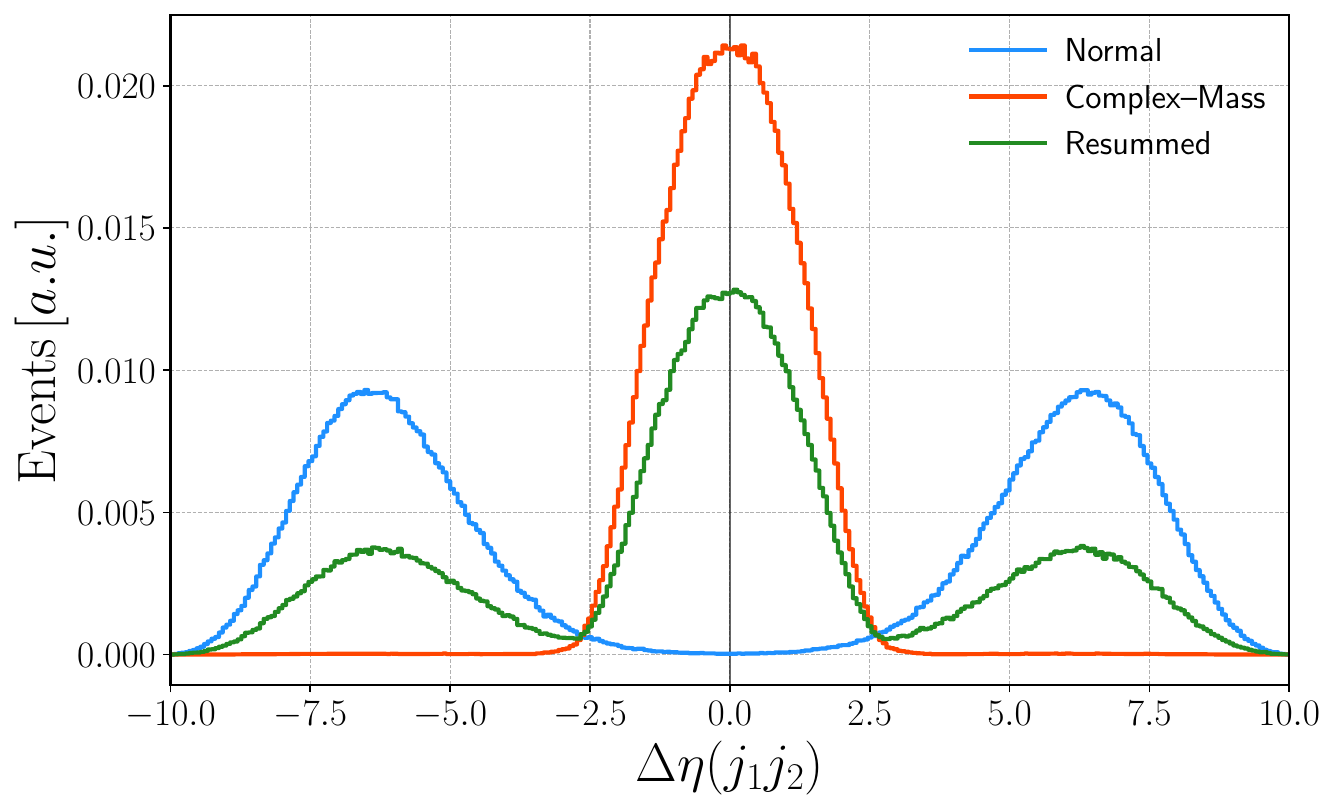}
    \caption{The $\Delta\eta(j_1 j_2)$ distributions (normalized to 1) for $pp\to\mu\nu_\mu jj$ at $m_N=10\,\mathrm{TeV}$ in the normal, complex-mass, and resummed schemes.}
    \label{fig:eta}
\end{figure}

\begin{figure}
    \includegraphics[keepaspectratio, width=1\linewidth]{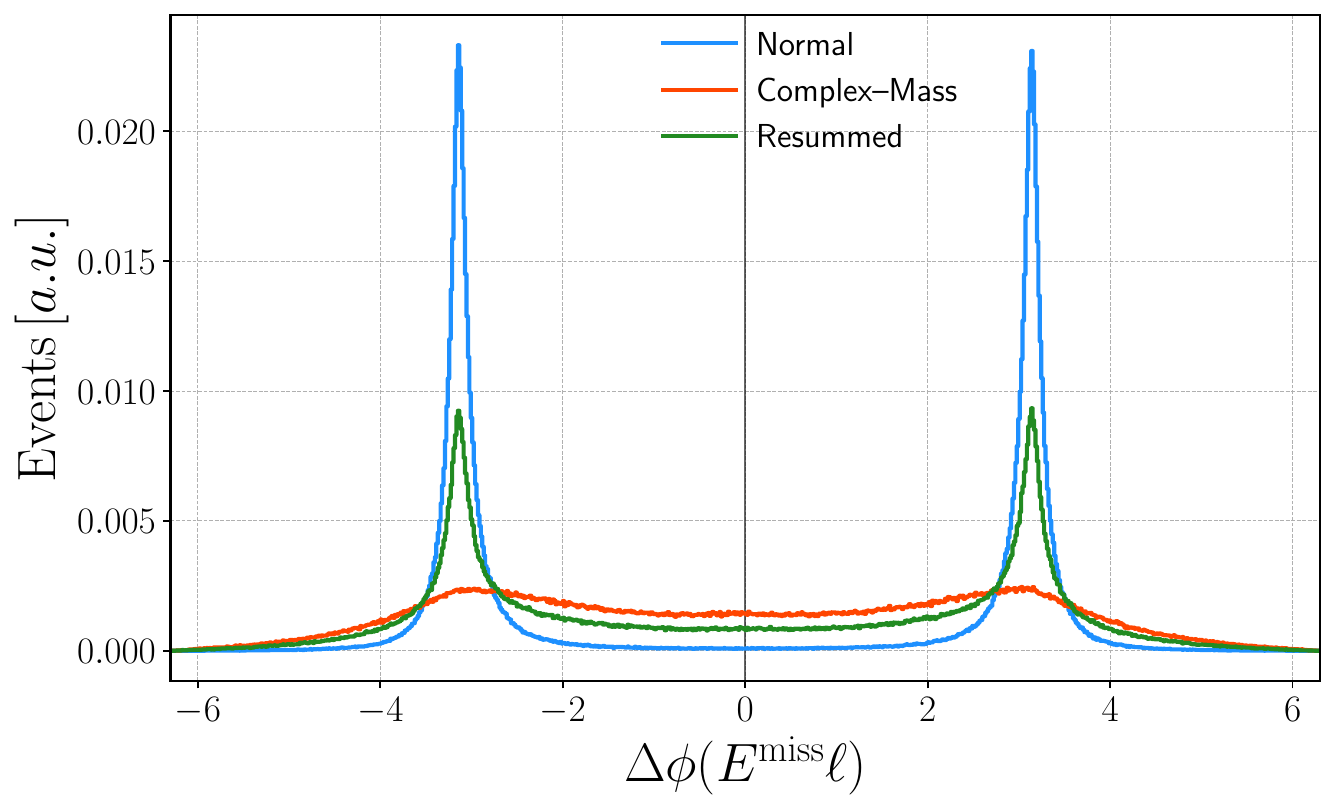}
    \caption{The $\Delta\phi(E^\mathrm{miss} \ell)$ distributions (normalized to 1) for $pp\to\mu\nu_\mu jj$ at $m_N=10\,\mathrm{TeV}$ in the normal, complex-mass, and resummed schemes.}
    \label{fig:phi}
\end{figure}

Given the rich kinematic structure of the signal, we illustrate the key differences among the three schemes using two complementary observables: the pseudorapidity separation of the jets, $\Delta\eta(j_1 j_2)$, and the azimuthal separation between the missing transverse momentum and the charged lepton, $\Delta\phi(E^\mathrm{miss}\ell)$, for $pp \to \mu \nu_\mu jj$ with a heavy-neutrino mass of $m_N = 10\,\mathrm{TeV}$. As shown in Figs.~\ref{fig:eta} and \ref{fig:phi}, the three schemes lead to distinct angular patterns. In $\Delta\eta(j_1 j_2)$, the normal scheme exhibits two broad peaks at large $|\Delta\eta|$, characteristic of the forward–backward jet configuration induced by $t$-channel exchange. By contrast, the complex-mass scheme produces a narrow and sharply central peak, reflecting an $s$-channel–like topology driven by an unphysically large Yukawa coupling. The resummed scheme interpolates between these extremes and consistently captures contributions from both channels; notably, even at $m_N = 10\,\mathrm{TeV}$, it retains a visible central component, contrary to the common expectation that $s$-channel effects should be negligible at very high masses.

A complementary picture emerges in $\Delta\phi(E^\mathrm{miss}\ell)$. The normal scheme exhibits strong peaks near $|\Delta\phi| \simeq \pi$, corresponding to a pronounced back-to-back configuration between the lepton and missing momentum—again consistent with a dominant $t$-channel topology. The complex-mass scheme instead yields a nearly flat distribution across the entire angular range, indicating the absence of such directional structure. The resummed result lies between these limits: while it maintains clear enhancements near $|\Delta\phi| \simeq \pi$, these peaks are noticeably softened relative to the normal scheme. Taken together, the two observables demonstrate that the resummed scheme preserves the physically relevant kinematic features of both channels while avoiding the unphysical distortions introduced by the complex-mass treatment.

We conclude this section by noting several directions for further study. First, while the resummed vertices satisfy the STIs for the longitudinal components, the transverse parts remain unconstrained; incorporating them—{\it e.g.}, to further improve ultraviolet behavior—would introduce additional model dependence and is therefore left for future work. Second, the substantial broadening of the propagator lineshape suggests that signal–background interference may become important, motivating a dedicated analysis beyond the scope of this work. Third, in the resummed scheme, we scale down the $V_{\ell N}=1$ benchmark to prevent the coupling from becoming arbitrarily large. However, the violation of perturbative unitarity simply indicates the breakdown of the perturbative expansion, not the breakdown of the underlying physics. As in low-energy QCD, an appropriate effective field theory could, in principle, be constructed to describe the dynamics of the strong-coupling regime. Finally, although we use the heavy neutrino model as a representative example, the same strategy applies to other broad resonances, including heavy bosons such as $Z^\prime$ and $W^\prime$~\cite{Pati:1974yy, Mohapatra:1974hk, Senjanovic:1975rk, CMS:2024vhy}, and could be incorporated into an automated patch for generating modified UFO models.\\
\indent \textit{Conclusion}---We have presented a practical and symmetry-consistent framework for incorporating large-width effects by combining a resummed heavy neutrino propagator with STI-implied resummed vertices, implemented directly in \textsc{MadGraph5}. Applying this approach to the Type-I seesaw model, we find that finite-width dynamics can qualitatively alter both the rate and kinematic structure of the signal. The propagator develops a sizable space-like component already at moderate widths, and the $s$-channel contribution remains relevant even at the multi-TeV scale. These effects distort the resonance lineshape, reshape angular correlations, and modify the scaling with $|V_{\ell N}|^2$, illustrating that default Breit–Wigner treatments can substantially overestimate collider sensitivities. As a result, existing searches for heavy neutrinos—and, more broadly, for any wide resonance—may require reinterpretation within a fully resummed framework.\\
\indent Although we focused on a single heavy neutrino as a representative example, the method applies to a wide class of models featuring broad states, including heavy $Z^\prime$ and $W^\prime$ bosons, and can be incorporated into automated UFO-generation workflows. Our results highlight that the large-width regime contains qualitatively new phenomenology that is not captured by standard approximations. This opens an opportunity for future experimental analyses, theoretical model building, and Monte-Carlo tool development to explore a region of parameter space that has been largely overlooked and where genuinely new signatures may emerge.\\
\indent \textit{Acknowledgements}---We gratefully acknowledge the support and funding provided by Vanderbilt University and the U.S. National Science Foundation. This work is supported in part by NSF Awards PHY-1945366 and PHY-2411502.

\bibliographystyle{apsrev4-2}

\end{document}